\documentclass[aps,prl,preprint,tightenlines,showpacs,byrevtex]{revtex4}
\usepackage{graphicx}

\begin{document}

\newcommand{\ra}{\rightarrow}
\newcommand{\calpha}{$\cos\alpha$}
\newcommand{\ctheta}{$\cos\theta$}
\newcommand{\BDPDstM  }{$B^0\ra D^+D^{*-}$}
\newcommand{\BDMDstP  }{$B^0\ra D^-D^{*+}$}
\newcommand{\BDPMDstMP}{$B^0\ra D^\pm D^{*\mp}$}
\newcommand{\BDSsPDstM}{$B^0\ra D^{(*)+}_s D^{*-}$}
\newcommand{\BDstDst  }{$B^0\ra D^{*+}D^{*-}$}
\newcommand{\pis }{$\pi_{slow}^-$}
\newcommand{\DP  }{$D^+$}
\newcommand{\DstM}{$D^{*-}$}
\newcommand{\BBb }{$B\overline B$}
\newcommand{\nDstDstnlsig}{$20\pm8$}
\newcommand{\nDsDstnlsig }{$27\pm5$}
\newcommand{\nDstDstlsig }{$3.0\pm1.2$}
\newcommand{\nDsDstlsig  }{$4.0\pm0.7$}
\newcommand{\Nnl    }{$283\pm103$}
\newcommand{\Nl     }{$37.6\pm13.2$}
\newcommand{\NfullE }{$28.0\pm7.0$}
\newcommand{\NfullM }{$29.6\pm6.6$}
\newcommand{\NfullBG}{$12.7\pm0.7$}
\newcommand{\effnl  }{$5.7\times10^{-3}$}
\newcommand{\effl   }{$8.4\times10^{-4}$}
\newcommand{\Brnl   }{$(1.57\pm0.57^{+0.30}_{-0.33})\times10^{-3}$}
\newcommand{\Brl    }{$(1.41\pm0.52^{+0.27}_{-0.30})\times10^{-3}$}
\newcommand{\Braverm}{$(1.48\pm0.38^{+0.28}_{-0.31})\times10^{-3}$}
\newcommand{\Brfull }{$(1.17\pm0.26^{+0.22}_{-0.25})\times10^{-3}$}

\vspace*{-3\baselineskip}
\resizebox{!}{2.5cm}{\includegraphics{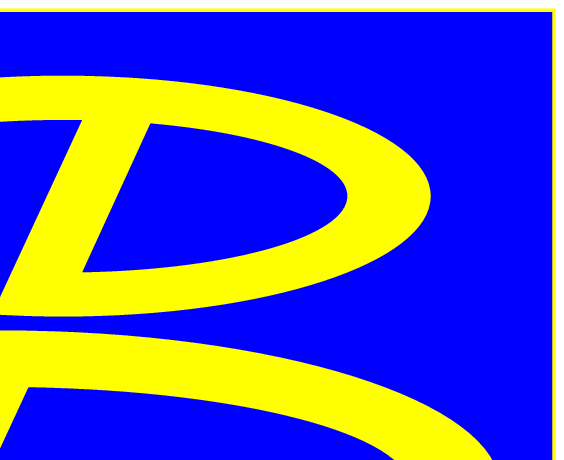}}
\preprint{\vbox{\hbox{KEK preprint   2002-42 }
                \hbox{Belle preprint 2002-16 } }}

\title{Observation of the decay $B^0\ra D^\pm D^{*\mp}$}

\author{
  K.~Abe$^{42}$,              
  R.~Abe$^{29}$,              
  T.~Abe$^{43}$,              
  H.~Aihara$^{44}$,           
  M.~Akatsu$^{22}$,           
  Y.~Asano$^{48}$,            
  V.~Aulchenko$^{1}$,         
  T.~Aushev$^{12}$,           
  A.~M.~Bakich$^{39}$,        
  Y.~Ban$^{33}$
% ...
\\      
\ \\
 (Belle Collaboration) \\
}

\affiliation{
$^{1}${Budker Institute of Nuclear Physics, Novosibirsk}\\
% ...
$^{12}${Institute for Theoretical and Experimental Physics, Moscow}\\
% ...
$^{22}${Nagoya University, Nagoya}\\
% ...
$^{29}${Niigata University, Niigata}\\
% ...
$^{33}${Peking University, Beijing}\\
% ...
$^{39}${University of Sydney, Sydney NSW}\\
% ...
$^{42}${Tohoku Gakuin University, Tagajo}\\
$^{43}${Tohoku University, Sendai}\\
$^{44}${University of Tokyo, Tokyo}\\
% ...
$^{48}${University of Tsukuba, Tsukuba}\\
% ...
}

\begin{abstract}
We report the first observation of the decay \BDPMDstMP\ with the
Belle detector at the KEKB $e^+e^-$ collider operated at the
$\Upsilon(4S)$ resonance.  The sum of branching fractions ${\cal
B}$(\BDPDstM)$+{\cal B}$(\BDMDstP) is measured to be \Brfull\ using
the full reconstruction method where both charmed mesons from $B^0$
decays are reconstructed. A consistent value (\Braverm) is obtained
using a partial reconstruction technique that only uses the slow pion
from the $D^{*-}\ra\bar D^0\pi^-$ decay and a fully reconstructed \DP\
to reconstruct the $B^0$.
\end{abstract}

\pacs{13.65.+i, 13.25.Gv, 14.40.Gx}
\maketitle


Mixing-induced $CP$-violating asymmetries have been recently observed
by the $B$ factory experiments using modes with charmonia in the final
state such as $B^0\ra J/\psi K_S^0$~\cite{gm}.  With little
theoretical ambiguity, the observed asymmetries determine $\sin2\phi_1$,
where $\phi_1$ is one of the angles of the unitary triangle of the
Cabibbo-Kobayashi-Maskawa matrix~\cite{ckm} in the framework of the Standard
Model.  Further understanding of $CP$-violation and probes of new
physics require measurements of $CP$-violation in additional decay
modes.  One class of such modes, the doubly-charmed decays $B^0\ra
D^{(*)}D^{(*)}$, has attracted a great deal of attention~\cite{bdd}.
While measurements of asymmetries in these modes are sensitive to
$\sin2\phi_1$, a deviation from the expected value is possible due to
penguin contributions with additional phases from new physics.  Among
the doubly-charmed decays, \BDPMDstMP\ is the most promising because
neither an isospin analysis nor an angular analysis is necessary to
measure $\phi_1$~\cite{bdsd}.  So far only an upper limit for the
branching fraction of \BDPMDstMP\ of $6.3\times10^{-4}$ was obtained
by the CLEO collaboration~\cite{cleofull}, although a na\"{\i}ve
expectation for this branching fraction is $1.0\times10^{-3}$, scaling
from the well measured branching fractions for the Cabibbo-allowed
processes $B^0\ra D^+_s D^{*-}$ and $B^0\ra D^-D^{*+}_s$~\cite{pdg}.


The data used for this analysis were taken with the Belle
detector~\cite{beldetec} at the KEKB asymmetric $e^+e^-$ ($3.5$ on
$8.0\,{\rm GeV}$) collider~\cite{KEKB}.  The integrated luminosity
used for this analysis is $29.4\,{\rm fb}^{-1}$ collected at the
$\Upsilon(4S)$ resonance ($\sqrt s=10.58\,{\rm GeV}$) (referred to
as on-resonance data) and $3.0\,{\rm fb}^{-1}$ at a
center-of-mass (CM) energy just below the threshold of \BBb\ pair
production (referred to as continuum data).  The on-resonance data
correspond to $31.9$ million \BBb\ pairs.  The Belle detector is a
large-solid-angle magnetic spectrometer that consists of a three-layer
silicon vertex detector (SVD), a 50-layer central drift chamber (CDC),
a mosaic of aerogel threshold \v{C}erenkov counters (ACC),
time-of-flight scintillation counters (TOF), and an array of CsI(Tl)
crystals (ECL) located inside a superconducting solenoid coil that
provides a $1.5$~T magnetic field.  An iron flux-return located
outside of the coil is instrumented to detect $K_L$ mesons and to
identify muons (KLM).


In the full reconstruction method we consider the decay \BDPDstM\
\cite{conjugate} followed by $D^{*-}\ra\bar D^0\pi^-$.  Five decay
modes are used for $\bar D^0$ reconstruction: $K^+\pi^-$,
$K^+\pi^-\pi^-\pi^+$, $K_S^0\pi^+\pi^-$, $K^+\pi^-\pi^0$ and $K^+K^-$;
\DP\ mesons are reconstructed through $K^-\pi^+\pi^+$ and $K_S^0\pi^+$
decay modes.

Charged tracks are required to be consistent with the hypothesis that
they originated from the interaction point in the $r$-$\phi$ plane.
Charged kaon candidates are required to be positively identified based
on the combined information from the ACC, TOF and CDC $dE/dx$ systems.
Only charged pion candidates involved in \DP\ reconstruction are
required to be positively identified to suppress the feed-down from
the Cabibbo-allowed $B^0\ra D^{(*)+}_s D^{*-}$ decays.

$K_S^0$ candidates are reconstructed from $\pi^+\pi^-$ pairs with a
common vertex that is displaced from the interaction point in the
$r-\phi$ plane and requiring an invariant mass within $\pm3\sigma$
($\pm7.2\,{\rm MeV}/c^2$) of the nominal $K_S^0$ mass.  $\pi^0$ meson
candidates are reconstructed from pairs of photons, each with energy
greater than $0.1\,{\rm GeV}$, that have an invariant mass within
$\pm3\sigma$ ($\pm13.5\,{\rm MeV}/c^2$) of the nominal $\pi^0$ mass.
The $\bar D^0$ and \DP\ candidates are required to have an invariant mass
within $\pm2\sigma$ of the nominal masses (the mass window depends on the
mode and varies from 8 to 22 ${\rm MeV}/c^2$),
where $\sigma$ is the
channel-dependent mass resolution.  \DstM\ candidates are formed from
a $\bar D^0$ and a slow pion candidate (\pis).  We require the mass
difference between \DstM\ and $\bar D^0$ to be within
$\pm11\sigma$($\sim5\,{\rm MeV}/c^2$) of the known mass difference 
in order to accommodate large non-Gaussian tails in the
$D^{*-}\ra\bar D^0\,$\pis\ mass distribution, which contain $16\%$ of the
signal.

\begin{figure}[tb]
\includegraphics[width=0.6\textwidth]{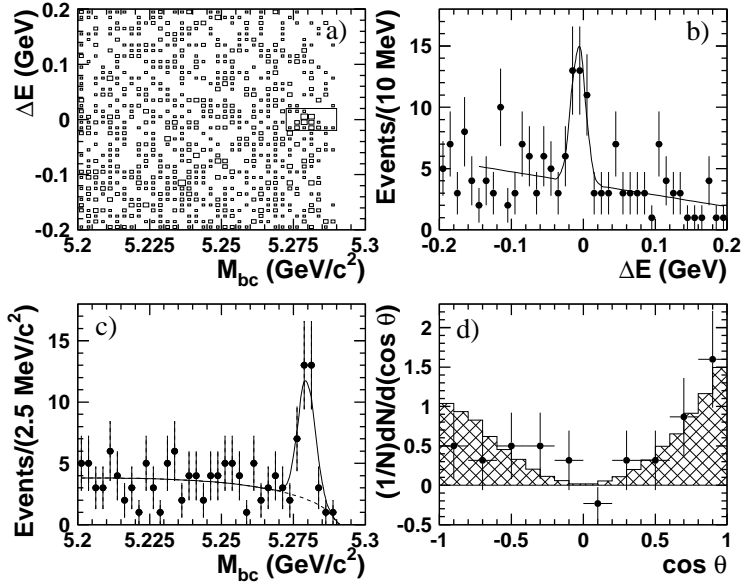}
\caption{Kinematical distributions of \BDPDstM\ candidates : 
a) scatter plot of $\Delta E$ versus $M_{bc}$; 
b) $\Delta E$ projection for $M_{bc}$ signal window, where the curve
represents the fit described in the text;
c) $M_{bc}$ projection for $\Delta E$ signal window, where the curve
represents the fit described in the text;
d) \ctheta\ distributions for the data (points with error bars) and 
for the MC signal (shaded histogram).}
\label{fig_scat}
\end{figure}

\BDPDstM\ candidates are selected using the CM energy difference,
$\Delta E\equiv E_B-E_{beam}^{CM}$, and the beam-constrained mass
$M_{bc}\equiv\sqrt{(E_{beam}^{CM})^2-P_B^2}$, where $E_B$($P_B$) is
the energy(momentum) of the $B$ candidate and $E_{beam}^{CM}$ is the
CM beam energy.  The scatter plot of $\Delta E$ versus $M_{bc}$, and
the projections onto $\Delta E$ and $M_{bc}$ are shown in
Fig.~\ref{fig_scat}{\it a,b} and {\it c}, respectively.  If more than
one candidate is found in an event, we use one of several criteria
(e.g. best vertex-constrained fit of the $D$ meson, number of CDC
hits, best $K_S^0$ or $\pi^0$ fit), depending on the kind of
ambiguity, to make a selection.  For example, the best
vertex-constrained fit is used to select among multiple $D$ meson
candidates.  We define the signal region by $|\Delta E|<0.02\,{\rm
GeV}$ and $5.2725<M_{bc}<5.2900\,{\rm GeV}/c^2$, corresponding to
$\sim\pm2\sigma$ and $\sim\pm2.5\sigma$ respectively.  Forty
candidates are found in the signal box.  Fig.~\ref{fig_scat}{\it d}
shows the background subtracted helicity angle ($\theta$) distribution
of the \DstM\ candidate. The distribution follows the $\cos^2\theta$
distribution that is expected for a pseudoscalar to pseudoscalar
vector decay.

The signal resolution in $M_{bc}$ is dominated by the beam energy
spread, while the $\Delta E$ resolution is dominated by the energy
resolutions of \DP\ and \DstM.  Therefore the $M_{bc}$ distribution is
less channel dependent and is more suitable for the extraction of the
signal yield.  We fit the $M_{bc}$ projection with a Gaussian and the
ARGUS background function~\cite{argfunc}.  The result is \NfullM\
signal events.  The statistical significance, defined as
$\sqrt{-2\ln({\cal L}(0)/{\cal L}_{max})}$, is $7.0\sigma$, where ${\cal
L}_{max}$(${\cal L}(0)$) is the maximized likelihood with(without) the
signal contribution.  A consistent yield (\NfullE) is obtained by
fitting the $\Delta E$ distribution with a Gaussian function over a
linear background.

As possible sources which can contribute to the signal peak, we
consider $B\ra D^+\pi^-_{slow}\bar D^0$ decay, where \DP\ and \pis\
come from resonant states ($D_1(2420)$ or $D_2(2460)$) or a
non-resonant state.  The resonant state contribution is checked using
the invariant mass of the \DP\ and \pis.  All the candidates have
invariant mass below $2.2\,{\rm GeV}/c^2$, hence there is no
indication of resonant states of \DP\ and \pis.  The non-resonant
contribution is checked using the Monte Carlo simulation (MC) and
found to be significantly suppressed by the \DstM\ selection
requirements.

The reconstruction efficiencies for each of the ten subchannels are
listed in Table~\ref{table_b_effi} (the MC statistical errors on the 
efficiencies are negligible).  For the branching fraction
calculation ${\cal B}(\Upsilon(4S)\ra
B^0\bar B^0)$ is assumed to be $0.5$.  The systematic error is
dominated by the uncertainty in the charged track reconstruction
efficiency ($2\%$ for each track).  The uncertainty in the signal
yield is estimated by varying the signal and background parameters in
the fit.  Contributions to the systematic error are summarized in
Table~\ref{syst}.  Finally, the branching fraction of \BDPMDstMP\ is
determined to be \Brfull, where the first error is statistical and the
second is systematic.
\begin{table}[tb]
\caption{The $B$ reconstruction efficiencies (\%) for the full
reconstruction.}
\begin{center}
\begin{tabular}{lccccc}
\hline
Channel & $K^+\pi^-$ & $K^+\pi^-\pi^-\pi^+$ & $K^0_S\pi^+\pi^-$
        & $K^+\pi^-\pi^0$ & $K^+K^-$ \\
\hline
$K^-\pi^+\pi^+$ & $10.7$ & $3.81$ & $3.88$ & $2.60$ & $7.78$ \\
$K^0_S\pi^+$    & $12.4$ & $4.39$ & $4.24$ & $2.89$ & $8.83$ \\
\hline
\end{tabular}
\end{center}
\label{table_b_effi}
\end{table}


We also use the \DstM\ partial reconstruction technique previously
used by the ARGUS and CLEO
collaborations~\cite{argpar,cleopar,cleodstpi} which benefits from a
higher $B^0$ reconstruction efficiency.  $B$ mesons produced in
$\Upsilon(4S)$ decays have low CM momenta ($\approx340\,{\rm MeV}/c$)
resulting in charmed mesons from \BDPDstM\ being almost back-to-back
in the CM frame.  The direction of the \pis\ from the \DstM\ decay
well approximates the direction of the mother particle due to the
small energy release in the \DstM\ decay.  Thus the CM angle $\alpha$
between \pis\ and \DP\ can be employed as a signature of this decay.
Another signature of the decay \BDPDstM\ is the \DstM\ polarization
angle ($\theta$) which can be calculated in the partial reconstruction
technique using kinematical constraints.  We use the CLEO definition
of the polarization angle in a partial
reconstruction~\cite{cleodstpi}:
\[
\cos\theta=\frac{-\beta_{D^*}(E^*_\pi-E^*_D)}{2P^*_\pi}
          +\frac{P_\pi^2-P^2_D}{2\beta_{D^*}\gamma^2_{D^*}M_{D^*}P^*_\pi},
\]
where $P_\pi$ ($P^*_\pi$) is the \pis\ momentum in the CM frame
(\DstM\ rest frame), and $E^*_\pi$ ($E^*_D$) is the energy of the
\pis\ ($\bar D^0$) in the \DstM\ rest frame.  $P_D$, $\beta_{D^*}$ and
$\gamma_{D^*}$ correspond to the $\bar D^0$ momentum in the CM frame
and the Lorentz factors for \DstM , calculated using energy
conservation and the known $B^0$ CM energy.  The polarization angle
distribution should exhibit a $\cos^2\theta$ behavior due to the zero
helicity state of the \DstM.

To suppress the backgrounds, which are much larger than for the full
reconstruction method, tighter \DP\ meson selection criteria are
applied.  \DP\ mesons are reconstructed using the $K^-\pi^+\pi^+$
decay mode only.  Tracks from \DP\ meson decays are required to have
associated SVD hits to guarantee a valid vertex reconstruction and are
refitted to a common vertex.  The good quality of the vertex fit is
required.  We exploit the relatively large \DP\ decay length and
require the decay path direction to be consistent with the momentum
direction of the \DP : $\cos(\vec V_D,\vec P_D)_{r-\phi}>0$, where
$\vec V_D$ is a vector from the interaction point to the reconstructed
\DP\ vertex.  This requirement has an efficiency of $87\%$ for real
\DP s and suppresses the combinatorial background by a factor of two.
The \DP\ CM momentum is required to lie in the interval
$1.63<P_{D^+}<1.97\,{\rm GeV}/c$ and the \pis\ is required to have CM
momentum smaller than $0.2\,{\rm GeV}/c$; these requirements are the
kinematic limits for the studied decays.

Some of the large backgrounds exhibit a similar angular tendency to
the signal, {\it e.g.}  the $c\bar c$ continuum background.  Continuum
events are partially removed by requiring the ratio of the second and
zeroth Fox-Wolfram moments~\cite{fox} to be smaller than $0.4$.  To
disentangle the signal and the continuum background, the whole data
sample is divided into two subsamples: the sample with a high-momentum
lepton in the event (referred to below as ``leptonic''), and the
sample without high-momentum leptons (``non-leptonic''sample).  Lepton
candidates are required to be positively identified as either a muon
or an electron and to have CM momenta larger than $1.1\,{\rm GeV}/c$.

In the leptonic data sample, the presence of a high-momentum lepton in
the event significantly reduces the continuum background.  This
background can be further suppressed using the angular correlation
between the lepton and the \DP\ candidate in the continuum.  We
require this angle to satisfy $-0.8<\cos(\vec P_{D^+},\vec
P_{\ell^\pm})<0.9$.  The continuum contribution is estimated to be
smaller than $1\%$ from the analysis of both continuum data and MC
simulation.  In addition, this requirement removes a significant
part of the \BBb\ background where the lepton and the \DP\ originate
from the same $B$ meson.

\begin{figure}[tb]
\includegraphics[width=0.6\textwidth]{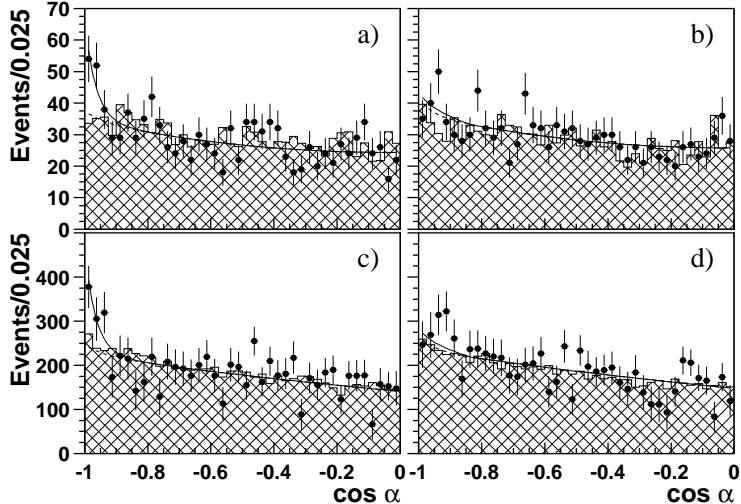}
\caption{a) and b) \calpha\ distributions for the leptonic sample, in
regions {\it A} and {\it B} respectively; c) and d) \calpha\
distributions after continuum subtraction in the non-leptonic sample.
Shaded histograms show the expected background distributions for
generic \BBb\ MC.  The curves represent the fit described in the
text.}
\label{dpnll}
\end{figure}

The selected \DP$\,$\pis\ combinations are divided into two
independent regions of the polarization angle.  The region
$0.5<|\cos\theta|<1.05$ (referred to below as region {\it {A}})
contains $85\%$ of the signal events, while the region
$|\cos\theta|<0.5$ (region {\it {B}}) is dominated by background.  The
distributions of \calpha\ for regions {\it A} and {\it B} are shown in
Fig.~\ref{dpnll}{\it a,b}. The points with error bars are the data
while the generic \BBb\ MC background distributions are superimposed
and shown as shaded histograms.  The \BDSsPDstM\ and \BDstDst\ decay
channels are excluded from the generic MC and are studied separately.

In the non-leptonic sample, the continuum data distributions of
\calpha\ are scaled by a factor to account for the difference in the
relative luminosities and cross-sections between on-resonance and
continuum data and are then subtracted from the on-resonance data.
The resulting \calpha\ distributions for regions {\it A} and {\it B}
are shown in Fig.~\ref{dpnll}{\it c,d}. The data are the points with
error bars while the generic \BBb\ MC background distributions are
superimposed and shown as shaded histograms.

The contamination from \BDSsPDstM\ and \BDstDst, which produce similar
peaks in the \calpha\ distribution are studied using the MC and
normalized to the branching fractions measured
in~\cite{cleopar,cleofull}.  We assume the \DstM\ is not polarized in
these channels; the effect of possible \DstM\ polarization is included
in the systematic error.  The contributions of \BDSsPDstM\ in region
{\it A} are estimated to be \nDsDstlsig\ and \nDsDstnlsig\ events for
leptonic and non-leptonic samples respectively; the \BDstDst\
contributions are \nDstDstlsig\ and \nDstDstnlsig\ events.  Several
other correlated backgrounds are studied in detail.  The
Cabibbo-allowed decays $B\ra D^{*-}D^{(*)+}K$ result in \DP\ CM
momentum below our requirement and are completely rejected by the
selection; for $B\ra D_1(2420)(D_2^*(2460))D^{(*)}$ decays the \DP\ CM
momentum is also shifted to lower values and these modes are strongly
suppressed.  The decays $B\ra D^{(*)-}\rho$ and $B\ra D^{(*)-}a_1$ are
studied with different polarization hypotheses and are found not to
produce a peak at \calpha$=-1$.  The possible contribution of these
background sources is included in the systematic error.

\begin{figure}[tb]
\includegraphics[width=0.6\textwidth]{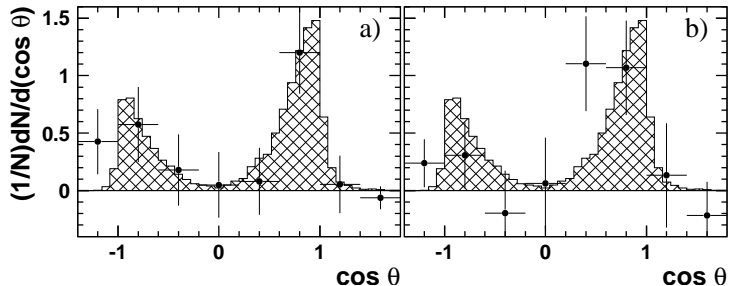}
\caption{Distributions of \ctheta\ : a) leptonic sample; b)
non-leptonic sample.  Shaded histograms show the signal MC.}
\label{polar}
\end{figure}

Finally, each pair of \calpha\ distributions, Fig.~\ref{dpnll}({\it
a,b}) and ({\it c,d}), is fitted simultaneously to signal and
background functions.  The signal is parameterized by an exponential
function which is a good description of the MC signal.  The ratio of
signal events in regions {\it A} and {\it B} is fixed according to the
MC.  The \BDSsPDstM\ and \BDstDst\ background contributions are fixed
in each region separately.  The shape of other combinatorial
backgrounds is parameterized by a second order polynomial that is
assumed to be the same for regions {\it A} and {\it B}, in agreement
with the MC.  The fit finds \Nl\ and \Nnl\ signal events in the
leptonic and non-leptonic samples, respectively.  As a cross-check, we
fit the \calpha\ distribution in eight bins of \ctheta . The resulting
polarization angle distributions are shown in Fig.~\ref{polar}{\it a}
and {\it b} for the leptonic and non-leptonic samples, respectively;
the expected $\cos^2\theta$ behavior can be seen in both samples.

One of the largest contributions to the systematic error in the signal
yield is the uncertainty of the background shape.  We use different
background parameterizations to check the stability of the result
against the choice of the background function.  We also perform a fit
with the background shape fixed from the generic \BBb\ MC.  A summary
of the systematic error sources is given in Table~\ref{syst}.
\begin{table}[tb]
\caption{Systematic errors on the branching fraction (\%).}
\begin{center}
\begin{tabular}{lcc}
\hline
Error source              &\hspace{1cm}Full          &\hspace{1cm}Partial       \\
\hline
Reconstruction efficiency &\hspace{1cm}$\pm17$       &\hspace{1cm}$\pm12$       \\
Particle identification   &\hspace{1cm}$\pm6$        &\hspace{1cm}$\pm4$        \\
Signal and 
background shape          &\hspace{1cm}$^{+1}_{-7}$  &\hspace{1cm}$^{+13}_{-15}$\\
Correlated background     &\hspace{1cm}   -          &\hspace{1cm}$^{+2}_{-3}$  \\
Number of \BBb\ pairs     &\hspace{1cm}$\pm1$        &\hspace{1cm}$\pm1$        \\
$D$ branching fractions   &\hspace{1cm}$\pm7$        &\hspace{1cm}$\pm7$        \\
\hline
Total                     &\hspace{1cm}$^{+19}_{-21}$&\hspace{1cm}$^{+19}_{-21}$\\
\hline
\end{tabular}
\end{center}
\label{syst}
\end{table}


The overall efficiencies estimated for the leptonic and non-leptonic
samples are equal to \effl\ and \effnl , respectively.  We calculate
the sum of branching fractions ${\cal B}$(\BDPDstM)$+{\cal B}$(\BDMDstP)
from the measurements with leptonic and non-leptonic samples to be
\Brl\ and \Brnl , respectively, where the first error is statistical
and the second is systematic.  Averaging these two measurements we
calculate ${\cal B}$(\BDPDstM)$+{\cal B}$(\BDMDstP)$=$\Braverm.


In summary, we report a measurement of the sum of the branching
fractions for the decays \BDPDstM\ and \BDMDstP\ using two analysis
methods.  The sum of the branching fractions ${\cal
B}$(\BDPDstM)$+{\cal B}$(\BDMDstP) is calculated to be \Brfull\ using
the full reconstruction method and \Braverm\ using the partial
reconstruction technique.  For calculations of the average branching
fraction in this channel, the result from full reconstruction should
be used.  The partial reconstruction result is a consistency check and
the sample obtained using this method will be used to improve
statistics in future measurements of $CP$-violation.  Our results are
higher than the upper limit from CLEO; they are however consistent
with the expectation based on $B^0\ra D^*D_s^{(*)}$.


We wish to thank the KEKB accelerator group for the excellent
operation of the KEKB accelerator.
We acknowledge support from the Ministry of Education,
Culture, Sports, Science, and Technology of Japan
and the Japan Society for the Promotion of Science;
the Australian Research Council
and the Australian Department of Industry, Science and Resources;
the National Science Foundation of China under contract No.~10175071;
the Department of Science and Technology of India;
the BK21 program of the Ministry of Education of Korea
and the CHEP SRC program of the Korea Science and Engineering Foundation;
the Polish State Committee for Scientific Research
under contract No.~2P03B 17017;
the Ministry of Science and Technology of the Russian Federation;
the Ministry of Education, Science and Sport of the Republic of Slovenia;
the National Science Council and the Ministry of Education of Taiwan;
and the U.S.\ Department of Energy.


\begin{thebibliography}{99}

\bibitem{gm}{
Belle Collaboration, K.~Abe {\it et al.},
Phys. Rev. Lett. {\bf 87}, 091802 (2001);
BaBar Collaboration, B.~Aubert {\it et al.},
Phys. Rev. Lett. {\bf 87}, 091801 (2001).}

\bibitem{ckm}{
M.~Kobayashi and T.~Maskawa, Prog. Theor. Phys. {\bf 49} 652 (1973).}

\bibitem{bdd}{
I.~Dunietz {\it et al.}, Phys. Rev. D{\bf 43}, 2193 (1991);
M.~Ciuchini {\it et al.}, Phys. Rev. Lett. {\bf 79}, 978 (1997);
A.~I.~Sanda and Z.-Z.~Xing, Phys. Rev. D{\bf 56}, 341 (1997);
X.-Y.~Pham and Z.-Z.~Xing, Phys. Lett. B{\bf 458}, 375 (1999).}

\bibitem{bdsd}{
Z.-Z.~Xing, Phys. Lett. B{\bf 443}, 365 (1998).}

\bibitem{cleofull}{
CLEO Collaboration, E.~Lipeles {\it et al.},
Phys. Rev. D{\bf 62}, 032005 (2000).}

\bibitem{pdg}{
D.~E.~Groom {\it et al.} (Particle Data Group), 
Eur. Phys. J. C{\bf 15}, 1 (2000).}

\bibitem{beldetec}{
Belle Collaboration, A.~Abashian {\it et al.},
Nucl. Instr. and Meth. A{\bf 479}, 117-232 (2002).}

\bibitem{KEKB}{
KEKB B Factory Design Report, KEK Report 95-1, 1995, unpublished.}

\bibitem{conjugate}{
The charge conjugate modes are implicitly included.}

\bibitem{argfunc}{
ARGUS collaboration, H.~Albrecht {\it et al.},
Phys. Lett. B{\bf 229}, 304 (1989).}

\bibitem{argpar}{
ARGUS Collaboration, H.~Albrecht {\it et al.},
Phys. Lett. B{\bf 324}, 249 (1994).}

\bibitem{cleopar}{
CLEO Collaboration, S.~Ahmed {\it et al.},
Phys. Rev. D{\bf 62}, 112003 (2000).}

\bibitem{cleodstpi}{
CLEO Collaboration, G.~Brandenburg {\it et al.},
Phys. Rev. Lett. {\bf 80}, 2762 (1998).}

\bibitem{fox}{
G.~C.~Fox and S.~Wolfram,
Phys. Rev. Lett. {\bf 41}, 1581 (1978).}

\end{thebibliography}
\end{document}